\documentclass[prb, reprint]{revtex4-1} 
\usepackage{amsmath}
\usepackage{soul}
\usepackage{amsfonts}
\usepackage{graphicx}
\usepackage{xcolor}
\usepackage{hyperref}
\usepackage{ulem}
\usepackage{url}
\usepackage{siunitx}

\begin{document}

\title{Spatial structuring of light for undergraduate laboratory}

\author{Sergei Panarin}
\author{Jonas M\"{u}ller}
\author{Shashi Prabhakar}
\author{Robert Fickler}
\email{robert.fickler@tuni.fi}
\affiliation{Photonics Laboratory, Physics Unit, Tampere University, Tampere, FI-33720, Finland}

\date{\today}

\begin{abstract}
In recent times, spatial light modulators have become a common tool in optics laboratories as well as industrial environment to shape the spatial structure of a beam.
Although these devices are often easy to use, they usually come at a high cost such that they are far from being implemented in a lot of undergraduate laboratories.
However, over the last years, the progress in developing more cost-effective projectors has led to affordable spatial light modulators in the form of so-called Digital Micromirror Devices (DMD).
This reduction in price, as well as their simple usage, make such devices increasingly suitable for undergraduate laboratories to demonstrate optical effects and the shaping of light fields.
Here, we show one of the most cost-effective ways to make a DMD available, namely turning a projector evaluation module into a computer-controlled spatial light modulator. We explain the underlying functioning and how this low-cost spatial light modulator can be used in undergraduate laboratories.
We further characterize the efficiency of the device for the most commonly used laser wavelengths and demonstrate various exemplary optics experiments suitable for undergraduate laboratories ranging from single and multi-slit diffraction, to optical Fourier transformations. 
Lastly, we show that using amplitude holography, the device can be used to generate transverse spatial modes, e.g. Laguerre-Gaussian beam, which are one of the most commonly used spatially structured beams.
\end{abstract}

\maketitle

\section{Introduction}
There is a lot of focus on the structuring of light to enhance various applications in the modern research fields, varying from microscopy to ultrafast pulse shaping to optical communications, optical trapping and quantum technologies. \cite{weiner_femtosecond_2000, dholakia2008optical, Willner15, Rubinsztein-Dunlop2017, erhard2018twisted}
All of these fields require methods to shape the transverse beam profiles using a so-called spatial light modulator (SLM).   
The SLMs themselves are devices that impose a transverse spatial modulation on a beam of light using either phase, amplitude, spatially varying polarization modulations, or their combinations.\cite{forbes2016creation, rosales-guzman_how_2017}
The great benefit of most of these devices is that they can be flexibly controlled by a computer using appropriate programs, such that their modulations can easily be adjusted to the specific field of applications or the task they perform.
These modulators are composed of 2D-array of micron-sized pixels so that the spatial modulations can be applied accurately with the precision of 8-10 microns.
Obviously, such devices would be very beneficial not only in research and industry, but also to demonstrate various optical effects and showcase simple applications using structured light.
Hence, SLMs are also a great tool in undergraduate student's laboratories for demonstrating various optical phenomena.\cite{padgett1996experiment, boruah2009dynamic, jones2016poincare, scholes_structured_2019}
However, many SLMs are too expensive for undergraduate laboratories to demonstrate simple experiments like studying diffraction and basic interference properties of a light beam.

To be able to make SLMs available to students and interested scientific enthusiasts on a frequent basis, more cost-friendly approaches are required.
While there are suggestions to utilize commercially available projectors as a way to implement a cost-effective solution to that problem,\cite{huang2012low} the described approach strongly relies on actual brand and, thus, on the availability of those devices.
In this work, we describe how to obtain and utilize a cost-effective and readily available SLM based on a digital micromirror device (DMD).\cite{Dmdappl1994} 
As DMDs are commonly used in cost-effective projectors, such as the projector evaluation module described here, they are often significantly cheaper than other SLM variants on the market.

We show how a very cost-effective evaluation module of a projector, including a DMD together with a single-board computer, can be turned into a versatile and convenient-to-use spatial light modulator. 
Our approach utilizes a combination of BeagleBone Black (BB)\cite{beaglebone} and DLP LightCrafter Display 2000 Evaluation Module (EVM),\cite{dlp2000EVM} both manufactured by Texas Instruments.
The BB is a single board computer and the EVM is a projector which contains an LED light engine, an optical system and, most importantly, the DMD model DLP2000. The latter which will be the essential part as it will be used as the spatial light modulator.
These devices combined provide a low-cost SLM (after the conversion described below) with a Linux-based platform, which usually costs less than 200 dollars.\cite{dlpprojector}
After a detailed description on how to turn the EVM into a fully computer-controlled SLM, we demonstrate a few experiments that could be used as showcases for the students' experiments ranging from diffraction, to Fourier optics and beam structuring.
Especially the latter might also be interesting for more advanced graduate courses\cite{galvez2006gaussian,boruah2009dynamic,jones2016poincare} and could also be utilized in demonstration experiments in outreach activities. 

\section{Implementing the DMD in undergraduate laboratories}
\begin{figure*}[!t]
    \includegraphics[width=14cm]{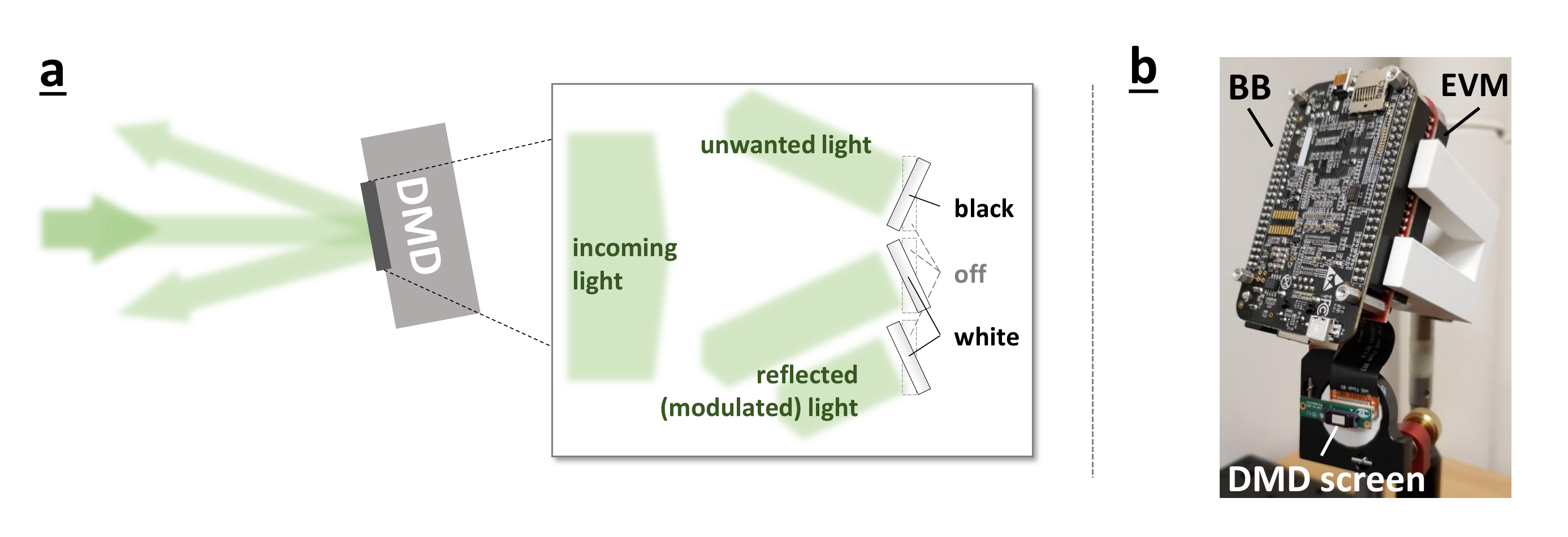}
    \caption{Functioning of DMD-based spatial light modulators. a) Modulating light with a DMD is implemented by reflecting light of a large array of small micro-mirrors, whose tilt is individually adjustable. Each mirror has three positions that depend on the device being switched off (gray dashed position in the background) or the color black or white is displayed. The DMD is placed in a setup in such a way that incoming light is only reflected into the direction of the optical setup by the pixels in the white position. Light that is reflected by mirrors in the black position (unwanted light) is redirected in a separate direction, i.e. neglected.
    b) Image of the bare DMD screen mounted on a mirror mount. The DMD was detached from the DLP DLCR2000 evaluation module (EVM) but is still connected to it, which is controlled by the BeagleBone (BB). The EVM and BB are mounted on top of each other and shown here mounted upside down in a 3D-printed holder.}
    \label{fig:dmd}
\end{figure*}

The projector evaluation module features the DLP2000 chip-set comprised of the DLP2000 0.2 nHD DMD, DLPC2607 display controller and DLPA1000 PMIC/LED driver.
The DLP2000 DMD has a resolution of 640$\times$360 pixels with 7.56 $\mu$m micromirror pitch, which is the separation between the centres of two consecutive micromirrors lying side-by-side.

There are several similar options available from Texas Instruments like EVMs based on the DMDs like DLP2010, DLP230GP, DLP230NP or others with a wide variety of specifications.
These products come at an extra cost but also provide more benefits, e.g. a better resolution or an easier control. 
Most of these devices, for example, are connected to the computer using a HDMI cable, and the computer recognises these devices as an extended display. 
Patterns displayed on the extended screen go directly to the DMD without the need for data transfer as it is the case for the presented cost-effective scheme.
However, as we focus on the most-cost effective realization of a spatial light modulator, we describe in detail, how the EVM DLP LightCrafter Display 2000 can be used as a simple, but versatile DMD.

The section is divided into three parts: understanding the underlying functioning of the micro-mirrors of a DMD, preparing the EVM device for optical experiments, and explaining its proper operations.

\subsection{Understanding DMD configurations}
The spatial light modulator we used, modulates the light by shaping its transverse amplitude, i.e. it enables to mask the light field by adjusting the direction of the reflected light with a pixel-to-pixel flexibility.  
This spatial-amplitude modulation is achieved by adjusting an array of micro-mirrors. Hence, its name digital micromirror device or DMD.
Here, each micron-sized mirror can be used to keep or remove the light's amplitude at a specific transverse position by reflecting it in or out of the direction of observation.
We note and demonstrate later that such an amplitude modulation also allows for phase modulation through amplitude holography, which will be discussed in section \ref{spatialmode}. 

The tilt angle between the off- and on-state is around 12\si{\degree}.\cite{dlp2000} 
In order to obtain the required modulations, the micro-mirrors of a DMD can usually be aligned in three different positions: i) off-state; ii) on-state corresponding to displaying black information; iii) on-state corresponding to white information (see inset in Fig. \ref{fig:dmd}a). 
We note that the micro-mirrors are not tilted along their edges but rather the diagonal of the mirror, which means that one corner of the mirror is lifted or lowered if the pixel is moved to the white or black position, respectively (see promotion video\cite{dlppicooverview} for more information). 
This diagonal tilt of the micro-mirrors results in a beam that is diagonally reflected up (or down) if the screen if mounted perpendicular to the optical table. 
Hence, we suggest to mount the bare DMD on an adjustable mirror mount (as can be seen in Fig. \ref{fig:dmd}b) such that this effect can be compensated and the reflected beam can be adjusted to the optical setup.

Although we used the device as a static spatial light modulator only slowly varying the displayed binary modulation patterns, it is worth noting that the DMD discussed in this article (similarly other devices) is mass-fabricated for the use in a projectors. 
In fact, we turned a projector EVM into a bare DMD to use it as a spatial light modulator. In projectors, the pixels are usually switched very fast, such that the three different encoding colors (red, green, blue) are modulated consecutively synchronized to different arrangements of the mirror array to form a proper color image.
In our system, a set of three LEDs of different colors is synchronized with the pixel-switching when the EVM is used in its original state, i.e. as a projector. 
Hence, after turning the EVM into a spatial light modulator, it should also have such fast switching-capabilities. 
While, we have not used nor studied this fast switching ability, it is worth noting that this modulation speed might be another interesting feature to test in undergraduate laboratories and has been used in research tasks with more advanced DMD models.\cite{mirhosseini2013rapid}

When a DMD is switched off but illuminated with a coherent laser field, there will be a diffraction pattern in the form of a 2D grid pattern visible.
This inevitable diffraction is due to the small space between the micro-mirrors pixels caused by the limited fill-factor (around 94\%\cite{dlp2000}) and reduces the overall efficiency of the device.
Another interesting fact, which relies on the working principle of a DMD, is that since the unwanted light is not absorbed but rather reflected into a different direction (see Fig. \ref{fig:dmd}a), the inversely modulated light field can be always observed too.
While the inverse amplitude can sometimes lead to completely different results,  for most holography it produces a similar or strongly correlated light field.
Finally, depending on the wavelength of the laser, one might also observe an additional reflection of the beam, which we attribute to stem from the screen surface window.

\subsection{Preparing the DMD}
The DLP lightcrafter is plugged into the BeagleBone black (BB) to get an embedded DLP projector device. 
Both, the DLP lightcrafter and the BB, should be powered by 5~V 1~A DC power source.
The BB should also be connected to the local area network (LAN) using a network cable for remote operations.
Here, we recommend starting the system first in its original state as a projector to verify its proper functioning and also to obtain the IP address. 
After the device boots up, a Debian desktop appears, which can be seen by projecting on a white screen, e.g. realized by a white paper or wall. 
It is also convenient to have a keyboard and a mouse connected to the BB via its USB port.
Once the projector is running, one can open the terminal from the menubar on bottom-left and enter the command \texttt{ifconfig}.
The line starting with \texttt{inet} contains the IP address of the board.
This address will be later required to access the board from a computer remotely.

After having tested the device as a projector, one can then turn the projector into an SLM. 
This procedure is simple, however, requires precise and careful handling of the device.
The essential step of this transformation is to separate the DMD DLP2000 from the optical engine and expose it for direct optical use.
This can be done by removing the screws, which secure the DMD and then separate the DMD by cutting through the glue with which it is attached to the optical engine.
Although the optical engine, i.e. the LEDs, are not required in the experiment, we do not recommend removing it from the board as this may stop the DMD from working due to a fail-safe configuration.
One can manually switch the LEDs off with a series of commands executed on the terminal of the BB (see Appendix \ref{ledoff} for more details).
If the device is not operated in a light-sensitive environment, having the LEDs working was very useful, as they served as a good indicator of the device being active. 
For light-sensitive applications, one can cover the LEDs with thick black paper or plastic sheet for a proper blockage of light. 

The final step is to mount the system for use in optical experiments.
For this, we glued the DMD screen onto a 3D-printed disk with 1 inch diameter to enable mounting into a commonly used mirror mount. 
For the control and driver unit, we also 3D-printed a holder, which allowed its flexible positioning on a breadboard or optical table (see Fig. \ref{fig:dmd}b). 
With these adjustments, we obtained a fully functional DMD as a simple but very versatile spatial light modulator, which can be given to undergraduate students for performing the experiment.

\subsection{Operating the DMD}
As the device modulates the amplitude in a binary fashion, i.e. ON and OFF, it is a best practise to use black and white (binary) images only. 
In general, the device's processor could be powerful enough, such that one can run a special program to prepare and display the required modulation patterns on the device itself, e.g. through a computationally effective program written in either Python or C/C++.
In our case, the binary images were produced using MATLAB on a standard desktop computer and transferred to the device (see appendix for details).

It is important to note that the device's processor has limited computational power. 
Sometimes a large image or having a few tasks running in the background can force the device to shut down. 
If this happens, it will take a few minutes to switch the board back on and resume the network connection.
Thus, we recommend testing the images to be displayed properly, before performing any experiment.

To transfer the generated binary images and display them as modulation patterns on the DMD, we used another computer on the network to connect to the BB using the program WinSCP which uses the SFTP protocol (see appendix).
To establish the connection to the DMD, the user enters the IP address of the device, obtained previously and then the username and password of the BB. 
The default username for a BeagleBoard is \texttt{debian} and the password is \texttt{temppwd}. 
The files can be transferred to an appropriate location on the storage.
The BeagleBone Black also has a micro-SD card slot, which can be used to increase the storage capacity of the device or prepare a repository of different modulation patterns beforehand using a separate computer.

To control the DMD, we further established a SSH connection through Putty (see appendix for more details), using the same IP address, username and password. 
After successful authentication, the BB is ready to execute commands for displaying the modulation patters on the DMD.
As the BB does not have an image viewer software installed, we chose \texttt{Feh} image viewer for its simplicity and lightweight (for installation, refer to appendix). 
The most useful commands to operate the device, i.e. to display the light-modulating patterns, are the following: 
\begin{itemize}
 	\item \texttt{export DISPLAY:=0}, a command that creates an environment variable \texttt{DISPLAY} that tells the device to display the image on the first display on local device, which is the DMD.
 	\item \texttt{feh -F image-name}, a command that tells the \texttt{Feh} viewer to display the desired image. \texttt{-F} suffix specifies it to open in full screen. \texttt{Feh} opens any format supported by imlib2, most notably jpeg, png, pnm, tiff and bmp.
 	\item \texttt{control+c}, is a combination of keys to close the \texttt{Feh} image viewer, which leaves the Debian desktop.
\end{itemize}
The DMD is now ready to be used as a programmable light modulator for spatially structuring a light field.

\section{Experiments for undergraduate students}

\subsection{Experimental setup for simple optics experiments}
\begin{figure}[htb]
    \centering
    \includegraphics[width=7.5cm]{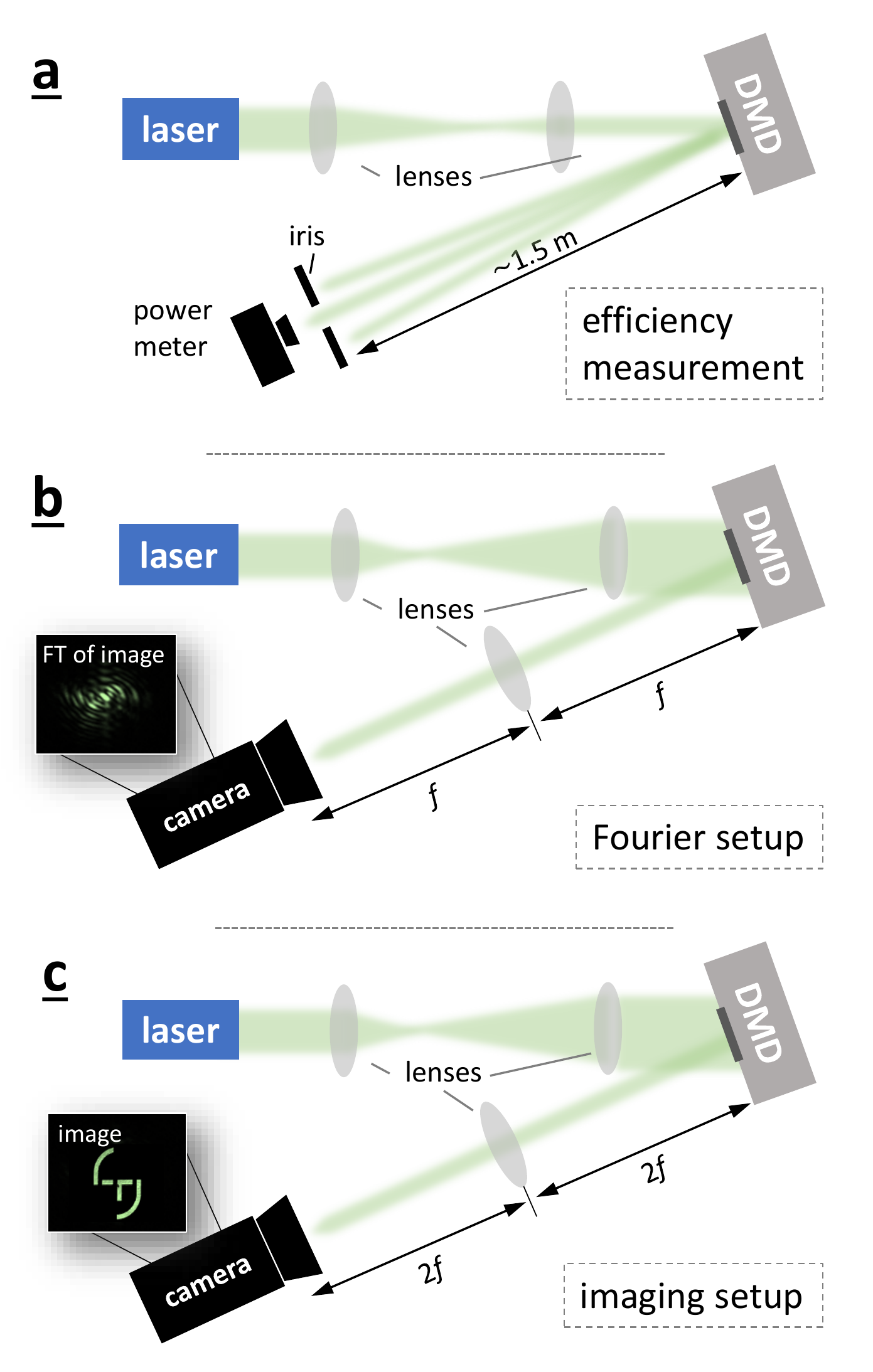}
    \caption{Three variations of the experimental setup. 
    a) Setup to measure the DMD's efficiency. Various laser beams of different wavelengths are adjusted in size by two lenses in a telescope arrangement to be smaller than the DMD screen. The reflected light is propagated over some distance, and its power in the different diffraction orders is filtered by an iris and measured using a power meter.
    a) 2$f$-Fourier setup. The laser beam is again adjusted in size by two lenses, this time to illuminate the entire DMD screen. A camera is placed in the direction of the modulated light and records the intensity structure. An optical Fourier transformation (FT) is performed by putting a lens between DMD and camera, each being one focal distance ($f$) away from the lens, and recording the intensity structure (inset shows an example). The setup was also used to observe the far-field intensity patterns of light diffracted through a slit of variable widths, multiple slits or forked gratings. 
    b) 2$f$-2$f$-imaging setup. The setup before the DMD is kept the same. However, after the reflection of the light, the lens is placed two focal distances (2$f$) away from the DMD, and the camera records the intensity structure after another two focal distances. Thus, the DMD screen is re-imaged on the camera.}
    \label{fig:setup}
\end{figure}

The ability to flexibly adjust the modulation using the DMD allows the performance of various experiments using nearly the same setup (see Fig. \ref{fig:setup} for the three variations of the setup we used).
For the first set of efficiency measurements, we adjusted the beam size of each utilized laser with an appropriate lens system using 2 lenses in a telescope arrangement (see Fig. \ref{fig:setup}a). 
We chose the lenses so that the beam diameter is smaller than the DMD screen, which ensures an accurate efficiency measurement.
The reflected beam was then propagated over a large enough distance (approx. 1.5 m) such that all diffraction orders were well separated and could be filtered out using an iris.
The obtained power in the different diffraction orders was measured with a power meter.
In the other two variations of the setup, the beam size of the laser (in these cases from a green laser diode) was adjusted using two lenses to make it larger than the DMD screen (see Fig. \ref{fig:setup}b and c), such that all pixels were illuminated with nearly equal intensity (plane wave illumination) and could be used for the modulation. 
Subsequently, the light was either recorded in the so-called Fourier plane, as shown in Fig. \ref{fig:setup}b or the image plane, as shown in Fig. \ref{fig:setup}c.
After reflection from the DMD, a recording of the intensity in the Fourier plane was achieved by placing a lens of focal length \textit{f} at a distance \textit{f} from the DMD and placing the camera at a distance \textit{f} from the lens. 
For measurements in the image plane, the distance between the DMD-Lens-Camera was changed to 2\textit{f} each.

We note that we provide a Matlab code, which generates the amplitude masks investigated in the subsequent experiments (apart from section \ref{sec:OFT}), as supplementary material.

\subsection{Measurement of DMD efficiency}  \label{efficiency}
Before we showcase a few exemplary experiments studying different suitable optical effects for undergraduate laboratories, we characterize the DMD in terms of its efficiency for different readily available laser wavelengths such as  405 nm, 532 nm, 650 nm, and 808 nm.

\begin{figure}[htb]
    \centering
    \includegraphics[width=8.5cm]{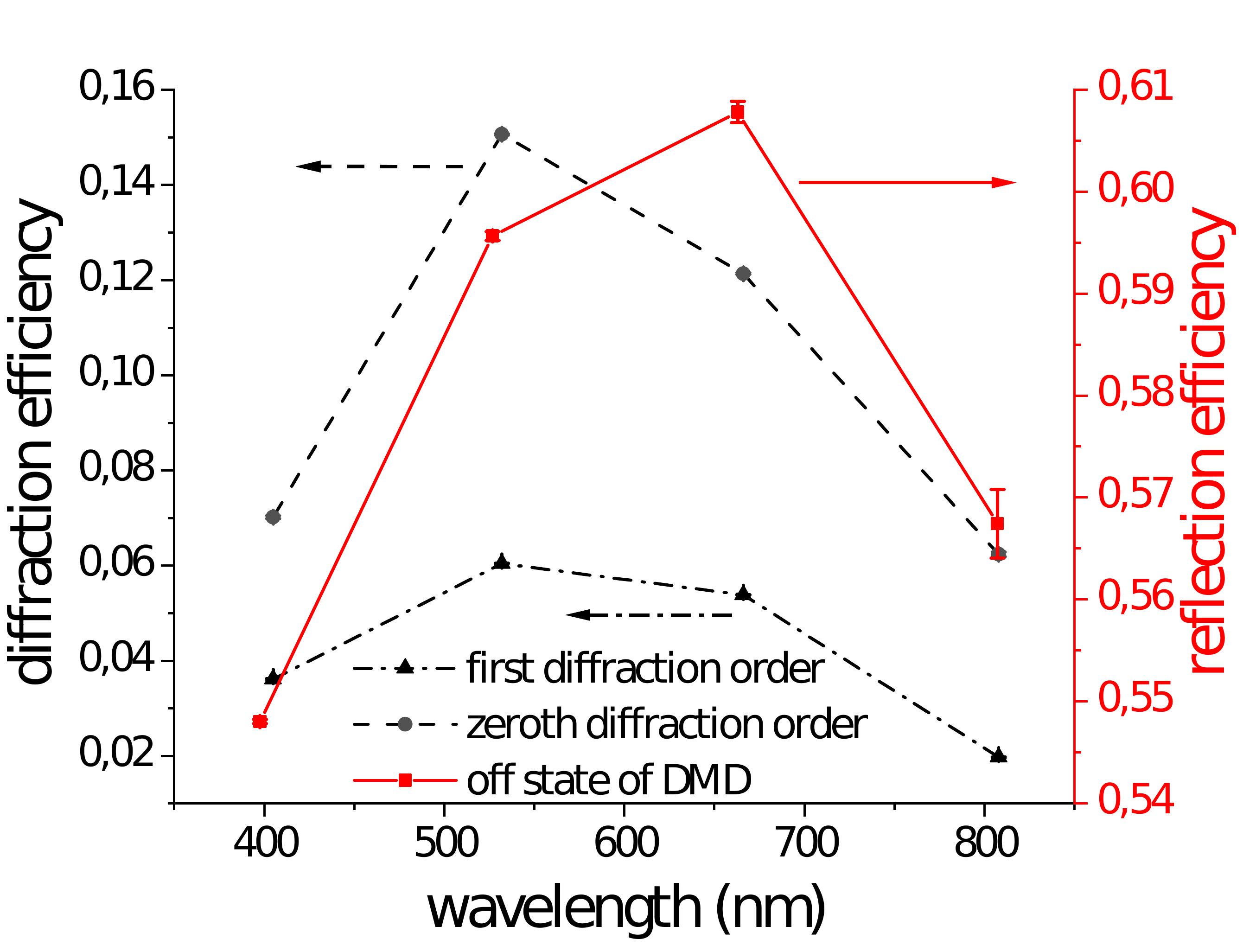}
    \caption{DMD reflection and diffraction efficiency varying with the wavelength of the laser light used. For most of the points, the error bars are too small to be visible. Arrows show the y-axis for corresponding plots. The efficiency has been found to be similar in positive as well as negative first order diffraction beams such that only one is plotted here. The type and color of the arrows match the corresponding lines.}
    \label{fig:efficiency}
\end{figure}

We performed and discuss the efficiency measurements for two reasons. 
First, we show that most of the available wavelength of cost-effective laser modules can be used and the DMD works similarly good. 
Second, a precise power measurement can also present various challenges for the students as it requires to isolate the desired light field from any external light sources as much as possible.
This was done in our case with an iris in front of the power meter to block the unwanted laser light and shield the detector of the power meter to ensure the best accuracy.
We note, that during this task, as well as all subsequent measurements, the use of a low power laser (eye safe, $<1 ~mW$) or a proper protection using laser safety googles is recommended.
Fine-tuning of the lenses' positions to adjust the beam size of the different lasers is also important in order to keep consistency between measured quantities of light reflected from the DMD and to make sure the entire beam hits the DMD.

In order to measure the reflection efficiency of the DMD, the micro-mirrors of the DMD must be unified into a single position.
We chose to use the off position, i.e. we performed these measurements with the SLM switched off. 
All losses found in this configuration can be attributed to absorption caused by an imperfect fill-factor of the DMD, i.e. gaps between the pixels, and the diffraction caused by the pixelization of the DMD screen.
The reflection efficiency in our experiment is then obtained as the ratio
\begin{equation}
    \epsilon=\frac{P_o}{P_i}, \label{eq:eff}
\end{equation}
where $P_i$ is the power of the input beam falling on the DMD, and $P_o$ is the measured output power after reflection. 

As expected for a device designed as a projector, we found that the reflection efficiency is slightly higher for light with a wavelength in the visible regime, i.e. 532~nm and 650~nm (see Fig. \ref{fig:efficiency} solid line and axis on the right hand side).
We found values of around 55-60~\% for all tested wavelength, which are in good agreement with the overall efficiency of around 66-67~\% specified by the manufacturer.\cite{dlp2000}

After these initial measurements, we then determined and compared the diffraction efficiency for the zeroth and first diffraction orders of a binary grating. The binary grating consisted of equally spaced on (white) and off (black) steps of 10 pixels (75.6~$\mu$m), which we displayed on the DMD. 
We determined the diffraction efficiency using the same formula as before \eqref{eq:eff}, this time with $P_o$ being the measured power in the targeted diffraction orders isolated by an iris and $P_i$ corresponding to the power of the incoming beam. 
The measured intensity in the first diffraction order is then determined by diffraction theory, which predicts a theoretically maximal diffraction efficiency of 10.1~\% for binary amplitude gratings,\cite{scholes_structured_2019, lee1979binary} which is further reduced by the aforementioned reflection efficiency. 
As required from the symmetry of binary gratings, we found very similar intensity values for both, the plus-first and minus-first diffraction orders.
We measured values between 2-6~\%, which also matches the theoretical value very nicely if the measured reflection efficiency is accounted for through simple multiplication of both percentages.
For the sake of clarity, we only show the measurements of the plus-first diffraction order in Fig. \ref{fig:efficiency} (dashed-dotted lines and axis on the left hand side).
For the zeros order, i.e. the light that is not diffracted by the binary grating and found in the center between the two first orders, we found around 6-16~\% (see Fig. \ref{fig:efficiency} dashed line and axis on the left hand side). 
From our measurements, we see that this DMD model has a very good efficiency at around 532~nm, which is a convenient to use wavelength for undergraduate laboratories as there are cheap laser modules available with eye-safe, yet perfectly visible power values.
We note that due to the tilt of the micro-mirrors, the DMD might also be used as a blazed grating if all mirrors are in the black or white color position. This configuration causes an angle depended diffraction as has been investigated in detail elsewhere,\cite{scholes_structured_2019} however, in this manuscript we focus on using the DMD as an amplitude modulating SLM, and all utilized gratings are binary amplitude gratings.

\begin{figure*}[htb]
    \centering
    \includegraphics[width=16cm]{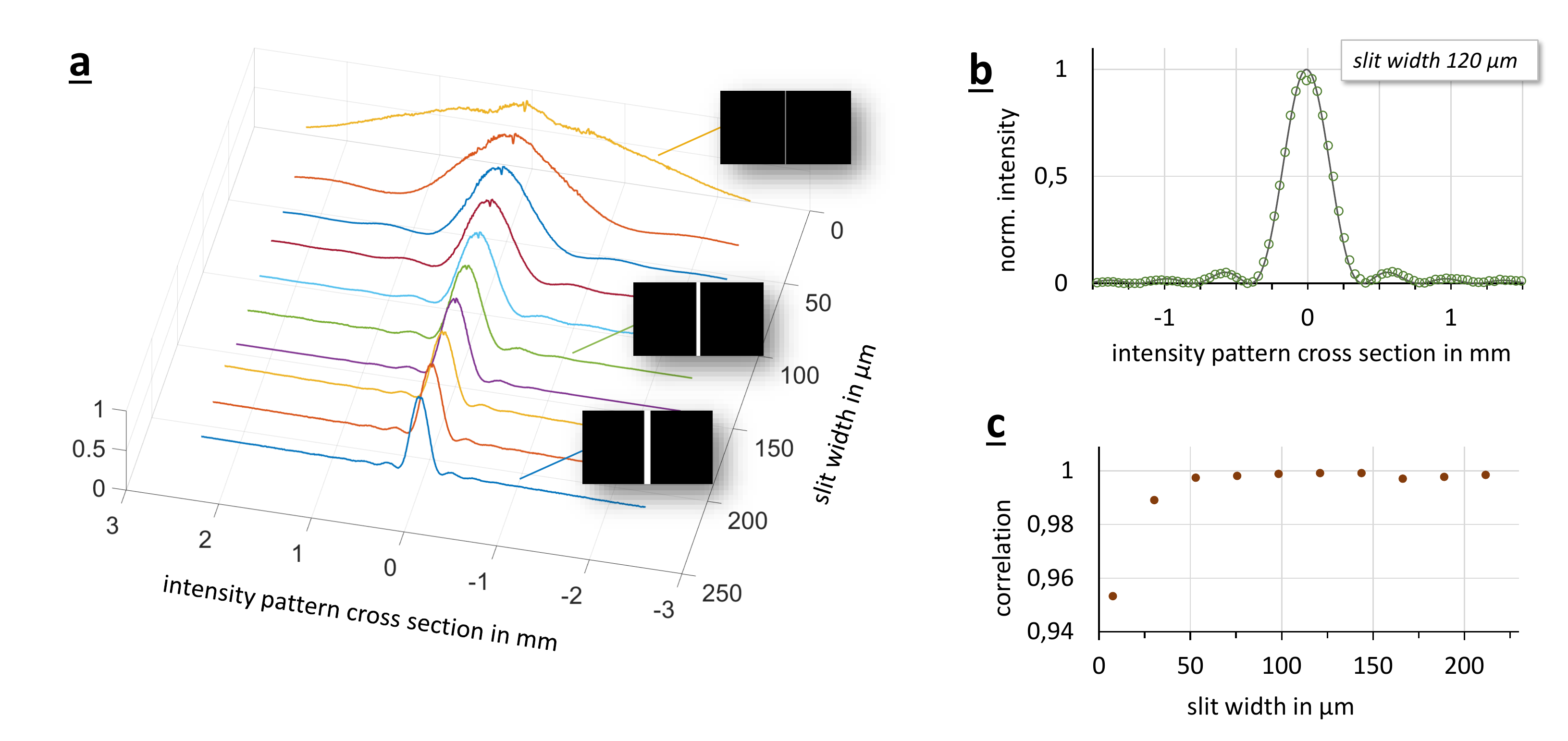}
    \caption{Intensity cross sections of diffracted light through slits of different widths. a) Decreasing the slit-width displayed on the DMD from 211.68 $\mu$m (28 pixels) to 7.56 $\mu$m (1 pixel) led to a broadening of the intensity pattern. The vertical axis represents the normalized intensity recorded using the camera. Three example slit-width patterns that were displayed on the DMD are shown as insets. b) Example plot of the measured data (green round circles) and the theoretically expected curve (black line, see also appendix D for more information), which has a near-perfect agreement, i.e. a correlation of 99,9\%. c) The quantitative analysis of the agreement between simulation and experiment for all cross-sections shows correlations of more than 99\% for nearly all measured patterns.}
    \label{fig:single-slit}
\end{figure*}

For the remaining set of experiments using the DMD, we used the green laser for which the DMD has the highest diffraction efficiency.
Moreover, this wavelength seems to be the best fit for undergraduate laboratories and demonstrations, as the human eyes also are the most sensitive in this regime such that eye-safe low-power laser modules can be used.

\subsection{Single-slit diffraction}

To showcase the advantage of using such a flexibly programmable SLM, we first investigated the diffraction of a laser beam through a single-slit of varying width. 
For this experiment, we used the setup in the Fourier configuration, i.e. the setup shown in Fig. \ref{fig:setup}b using a lens with the focal length of 15~cm. 
To demonstrate the evolution of the intensity pattern obtained from a single-slit diffraction of varying width, we changed the slit widths from 28 pixels (211.68~$\mu$m) to 1 pixel (7.56~$\mu$m). 
This evolution of the diffraction pattern is shown in Fig. \ref{fig:single-slit}a, where we plot the intensity profiles for ten different slit widths. 
One nicely observes the transformation of the slit shadow image for a very wide slit to a widely diffracted profile obtained from diffraction through very narrow slits.

To test the quality of the patterns obtained by a single-slit diffraction, we compared the obtained pattern with the theoretically simulated ones. 
We simulated the theoretically expected patterns using a split-step beam propagation method\cite{poon2017engineering} using the same physical values, i.e. wavelength, propagation distances, lens etc., as the ones in our experiment.
As an example, we plot the line profile of the pattern obtained for a slit width of 120~$\mu$m in Fig. \ref{fig:single-slit}b. 
The line shows the theoretically calculated distribution and the points represent the experimental data points. 
One can see that the data fits very well to the theory. 
To further quantify our result, we evaluated the correlation function\cite{lee1988thirteen}, which is a measure of how close the two data sets match (see appendix for more details). 
All our obtained data agrees very well with the theory, i.e. we obtained correlation values with more than 99\% for nearly all measurements. 
Only for the arrangement with the narrowest slit, the correlation drops to 95\% as can be seen in Fig. \ref{fig:single-slit}c. 
This quantification affirms the accuracy of the experimentally measured diffraction performed by the cost-effective DMD.

\subsection{Multiple-slit diffraction}
\begin{figure*}[!t]
    \centering
    \includegraphics[width=16cm]{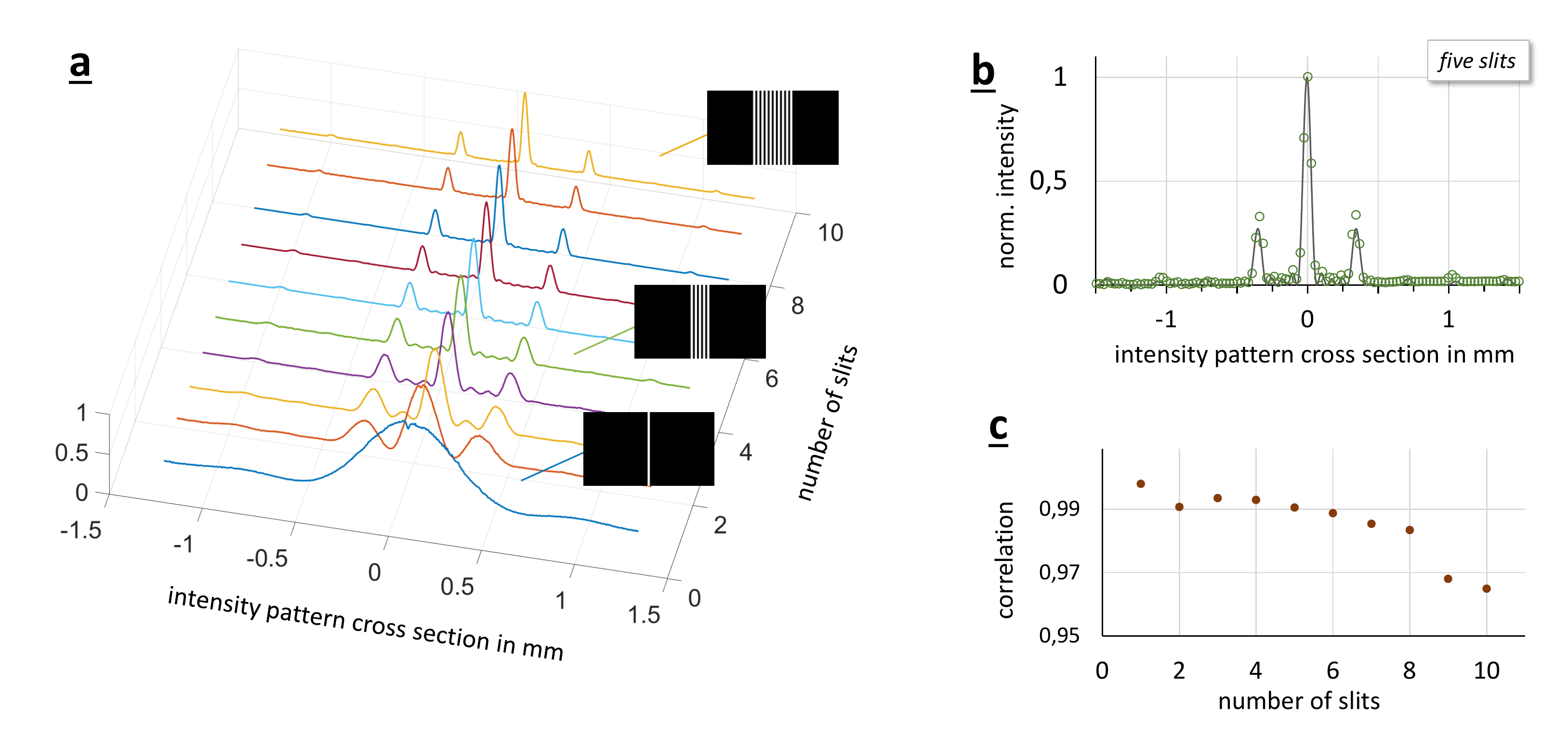}
    \caption{Intensity cross sections of diffracted light through various number of slits. a) By increasing the number of slits from 1 to 10, one can obtain a continuous change from the diffraction through one or a few slits to a diffraction through a grating. The width of the slits and spacing between the slits was kept constant to be 10 pixels, i.e. 75.6~$\mu$m. The vertical axis represents the normalized intensity recorded using the camera, and the insets show three example patterns for one, five and ten slits. b) Comparison between theoretically expected curve (black line, see also appendix D for more information) and the measured data (green round circles). The agreement is again very good, with the two side lobes being slightly too high in the measured intensity. Nevertheless, the displayed graph in b) as well as the quantitative analysis of all measurements depicted in c) show very high correlations of more than 99\%, only slightly decreasing to around 96\% for 10 slits.}
    \label{fig:multiple-slit}
\end{figure*}

As a second experiment investigating the wave nature of light, we studied the diffraction of light through an increasing number of slits.
Again, the Fourier setup shown in Fig. \ref{fig:setup}b is used, this time using a lens with a focal length of 20~cm. 
We started with a single slit of ten pixels width (75.6~$\mu$m) and then increased the number of slits by one for each recording (distance between the slits is also set to 10 pixels). 
In the final arrangement due to a large number of slits, the diffraction can already be considered equal to that of a grating, with two dominant, symmetric diffraction orders.
We recorded images for each setting and plot the intensity profiles across the maximum intensity in one graph (see Fig. \ref{fig:multiple-slit}b). 
From these measurements, one can nicely observe the relatively smooth transition from a single slit diffraction pattern to that of a double, triple and multiple-slit pattern. 
The higher the number of slits, the clearer the diffraction pattern of a binary grating appears.
Similar as before, we simulated the theoretically expected intensity cross section and compared it to our experimentally recorded one.
In Fig. \ref{fig:multiple-slit}b, we show how well the experiment (data points) matches the simulation (line) for a five-slit diffraction pattern. 
Evaluating the correlation for all recorded patterns (see Fig. \ref{fig:multiple-slit}c) shows again a very good agreement of more than 99\%, with a slight decrease to 96\% for a larger number of slits.

Thus, the demonstrated set of measurements and its comparison with the theory can be used in undergraduate optics laboratories to show the effect of a varying slit width as well as the number of slits with respect to the diffraction of light. 
Although we have used a split-step propagation method to simulate our experiment theoretically, the results are also related to an optical Fourier transformation, which we will explore in the following subsection in more detail.

\subsection{Optical Fourier transformation}\label{sec:OFT}
\begin{figure}[!b]
    \includegraphics[width=8.5cm]{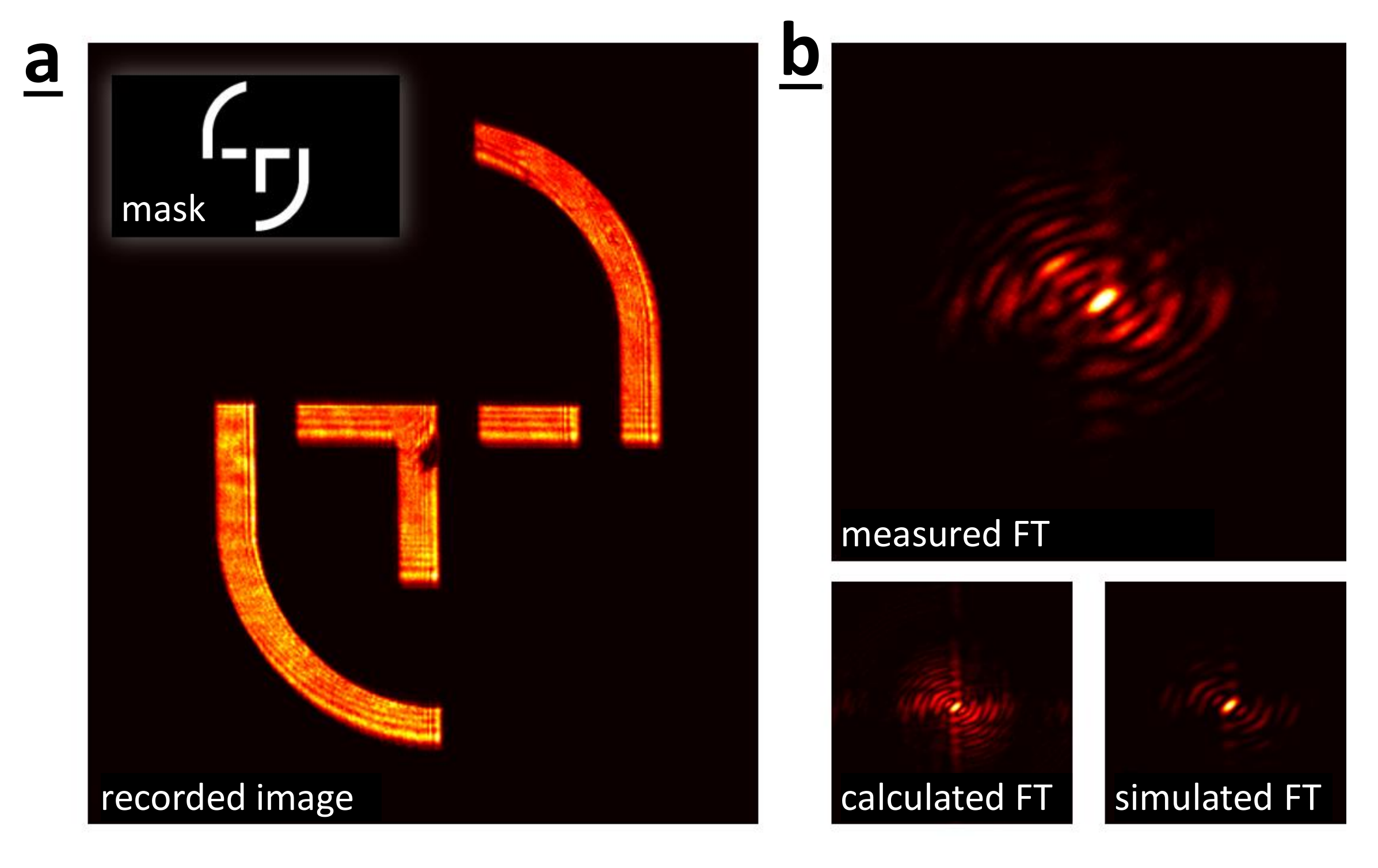}
    \caption{Imaging and optical Fourier transformation using the DMD. a) Using a mask (here the Tamper University logo, see inset) and a 2f-2f imaging system, one can generate an image using the DMD. b) When implementing a lens in a 2-f configuration, one can perform an optical Fourier transformation (FT) of the image. The upper image shows the recorded intensity pattern, which nicely matches a calculated Fourier transform (lower left side) as well as the simulation of the experimental setup performing an optical FT (lower right side).A quantitative comparison results in a correlation of 88\% or 85\% between the recorded image and the calculate or simulated result, respectively.}
    \label{fig:TUNI}
\end{figure}

Another possible use for the presented cost-effective DMD in undergraduate laboratories is to demonstrate an optical Fourier transform (FT). 
To showcase the implementation of optical FT, we used the logo of our university, Tampere University. 
To obtain a clear intensity image of the logo displayed on the DMD, we used the imaging configuration shown in Fig. \ref{fig:setup}c with a lens of 20~cm focal length.  
The utilized mask of the logo and the recorded image are shown in Fig. \ref{fig:TUNI}a. 
After this initial test, we then placed the lens and camera back to the positions preforming the Fourier transformation (Fig. \ref{fig:setup}b).
With this configuration, the recorded intensity structure of the light corresponds to the Fourier transform of the original image (more details on the theoretical correspondence can be found in standard text books\cite{goodman2005introduction, hecht2002optics}).

To verify the correctness of the obtained intensity pattern, we compare our measured FT with theoretical predictions. 
On the one hand, we compared it to use a 2D fast Fourier transformation, e.g. using Matlab, and on the other hand to the simulated intensity profile using the split-step beam propagation method as before (see Fig. \ref{fig:TUNI}b).
Again, we evaluated the correlation between our measurement result and the calculated FT as well as the simulated one. We got an agreement of 88\% and 85\%, respectively, which shows that that indeed light passing through lens results in an intensity pattern corresponding to the Fourier transform.
Thus from these comparisons, we found that our setup works very well with only minor imperfection leading to slight discrepancies, which are expected due to the low-cost implementation.

After having investigated the possibilities to study optical Fourier transformations, we continue to give simple examples of how amplitude modulations can result in beams with varying phase fronts.

\subsection{Fresnel Zone plate}
At first, we tested the focusing effect using only amplitude modulations, i.e. we implemented a so-called Fresnel zone plate (FZP). 
For this, we created the pattern of a FZP with a focal length of 10 cm, display it on the DMD, removed the lens after the DMD from the optical setup, and kept the plane wave illumination. 
We found that the beam focuses to a small spot after propagated to the focal plane.
\begin{figure}[h]
    \centering
    \includegraphics[width=8cm]{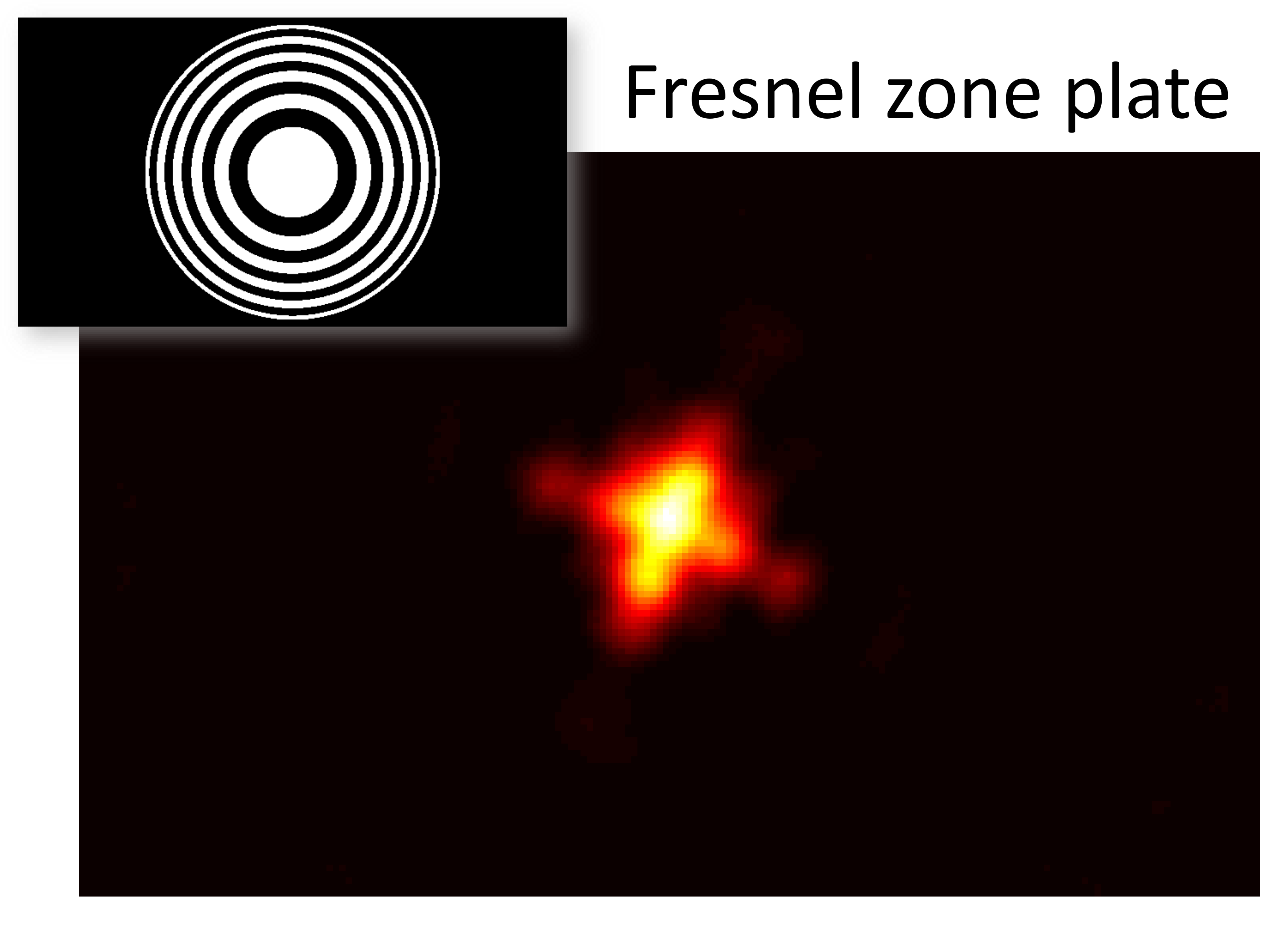}
    \caption{Focusing using a Fresnel zone plate. The inset shows the amplitude used in the experiment. The focus recorded by the camera deviates from a Gaussian spot, i.e. it is found to be distorted due to astigmatism caused by the DMD.}
    \label{fig:fresnel-zone}
\end{figure}
When placing our camera at that plane, however, we not only saw that the beam is tightly focused to a Gaussian spot but also slightly deformed (see Fig \ref{fig:fresnel-zone}).
Because we ensured the circular symmetry of the cutout beam at the DMD by displaying the FZP only at a circular region, we relate this deformation to astigmatism of the DMD.
This might stem from the imperfect optical flatness of the device due to the rather low production costs and can also be seen for other spatial light modulators.\cite{jesacher2007wavefront,scholes_structured_2019} 
We further confirmed astigmatism in the following experiment, in which we show how to holographically generate spatial light modes through specialized gratings.

\subsection{Generation of spatial modes using holography} \label{spatialmode}
As the last demonstration, we utilized the DMD DLP2000 to generate transverse spatial modes, in particular so-called Laguerre-Gaussian (LG) modes or donut beams. 
These beams have attracted a large amount of attention over the last decades in various fields due to various interesting properties attached to the spatial structure.\cite{allen1992orbital, erhard2018twisted, Willner15}
The main feature of LG beams is their screw-like phase structure, i.e. a twisted helical wavefront of the form $e^{i \ell\phi}$, where $\phi$ is the azimuthal coordinate and $\ell$ corresponds to the quantized integer orbital angular momentum (OAM) value,\cite{allen1992orbital} which can take on values as large as 10,010.\cite{Fickler2016}
Because of this twisted structure, the beams have a phase singularity along the optical axis, which then leads to an intensity null and, thus, a donut-like intensity pattern.\cite{allen1992orbital}

Since our DMD is only able to modulate the amplitude, we had to use holography techniques to imprint such helical phase structures, thereby, generating donut beams.
These holograms correspond to diffraction gratings with dislocation at the optical axis, which resembles a fork structure, hence, the name forked gratings.\cite{galvez2006gaussian, carpentier_making_2008, Lerner:12}
Spatial light modulators have been extensively used to generate the OAM modes \cite{galvez2006gaussian, boruah2009dynamic, jones2016poincare, scholes_structured_2019} using forked gratings.
Depending on the strength of the dislocation one generates donut beams with different topological charges, i.e. helical wavefronts for which the transverse phase increases along the azimuthal direction in multiples of 2$\pi$.

\begin{figure*}[!t]
    \centering
    \includegraphics[width=17cm]{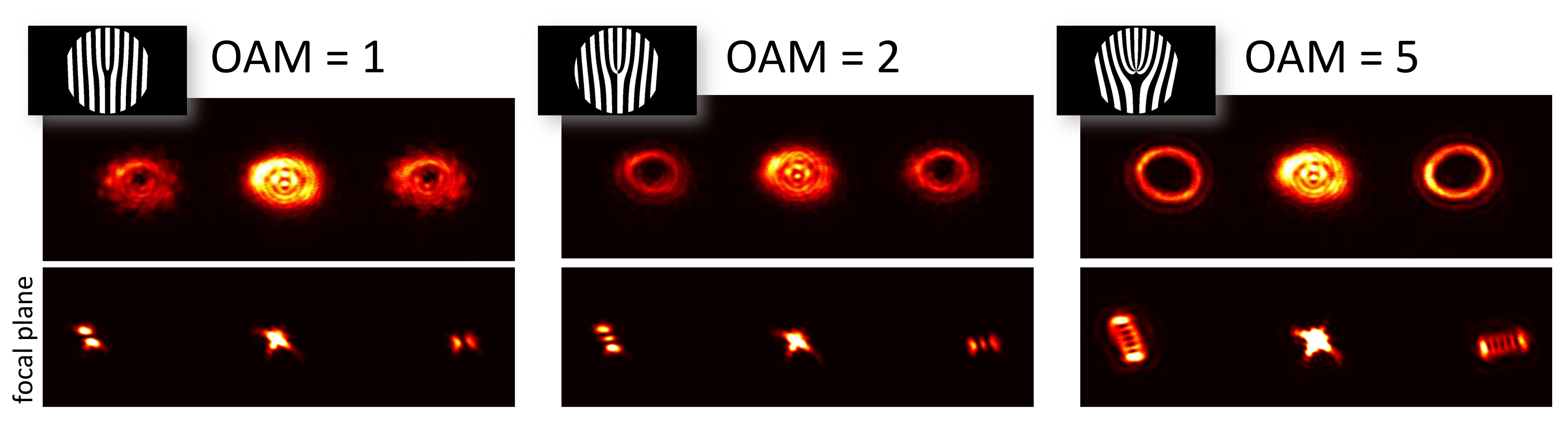}
    \caption{Beam shaping via forked diffraction gratings. We generated three different donut beams carrying 1 (left), 2 (center) and 5 (right) quanta of OAM. The insets show the utilized forked diffraction gratings, which led to donut shaped beams (upper row) in the both first diffraction orders. The lower row shows the intensities in the focal plane, where the beams get distorted heavily due to astigmatism caused by the DMD such that Hermite-Gaussian beam patterns appear.\cite{beijersbergen_astigmatic_1993}}
    \label{fig:OAM}
\end{figure*}

In our experiment, we used fork diffraction gratings of the topological charge of 1, 2 and 5 (see insets in Fig \ref{fig:OAM}), which leads to donut beams carrying 1$\hbar$, 2$\hbar$ and 5$\hbar$ of OAM per photon.
To better observe the structure of modulated beams, we again implemented after the DMD a lens of focal length of 25~cm in 2f configuration, as shown in Fig. \ref{fig:setup}b.
When putting the camera close but not exactly into the focal plane (approximately 5~cm away), we observed the expected donut-like intensity pattern in both first diffraction orders. 
We also saw that the beam diameter increases when larger topological charges are imprinted, leading to an increased divergence of the beam\cite{curtis2003structure} (see Fig \ref{fig:OAM} upper row).
If we put the camera in the focal plane, we expected the same mode structure to appear. However, due to the above-described astigmatism, we found that the structure resembles another well-known spatial mode, i.e. a Hermite-Gaussian (HG) mode (see Fig \ref{fig:OAM} lower row).
This can be explained by the fact that astigmatism resembles a focusing using a cylindrical lens, which is known to transform LG modes into HG modes.\cite{beijersbergen_astigmatic_1993}

This observation confirmed that the DMD causes astigmatism when modulating the light. 
Although it would be possible to correct for the astigmatism and aberrations using the obtained images,\cite{jesacher2007wavefront,scholes_structured_2019} we instead took advantage of this effect and used it to verify the imprinted topological charge. By simply counting the number of observed lobes and their orientation, we could verify the OAM quanta and its sign.\cite{beijersbergen_astigmatic_1993,padgett1996experiment} 
We observed the same amount of lobes for both first diffraction orders and that the number of lobes agrees with the topological charge that we set out to imprint onto the beams, i.e. the OAM value minus one.
We further saw that the observed patterns in both first orders are orthogonally oriented, which proves that both beams are carrying OAM values of opposite signs. 
Thus, we have verified that using the DMD enables the generation of transverse spatial modes of light, such as the donut-shaped LG modes.

\section{Conclusion}
The use of a BeagleBone device together with DLP lightcrafter evaluation module is a very user-friendly device and money-saving technique of obtaining an SLM in the form of a DMD. Its quality is good enough for various educational purposes, e.g. their use in undergraduate laboratories for demonstrating instructive experiments.
In this manuscript, we have described in detail how the evaluation module, designed to be a projector, can be turned into a DMD, explained its underlying functioning and appropriate usage.
We have further presented a set exemplary experiments that provide a gradual increase in complexity, ranging from diffraction to Fourier optics and beam shaping, while at the same time retaining the easiness of realization. 
Simple experiments, like the ones presented here, can communicate necessary knowledge to students without overloading them with too much information and complicated procedures, which will help students (or even interested laymen) to better grasp the investigated optical effects. 
With our setup, students will also be able to gain firsthand experience with simple programming as well as modern electronic devices such as projectors using spatial light modulators. 
Although we described the procedure used to modify one particular model, other LightCrafter models from Texas Instruments can also be converted using the same process.
The presented scheme will make spatial light modulators available in a very cost-effective way such that they can be easily employed in undergraduate as well as undergraduate laboratories.
Their use will not only facilitate demonstrations of a broad range of optical effects but also enable students to study themselves various optical effects and applications like interference, phase shaping, holography, beam steering, to name a few.

\begin{acknowledgments}
The authors thank Markus Hiekkamäki and Lea Kopf for helpful discussions. SP, JM, SP and RF acknowledge the support from Academy of Finland through the Competitive Funding to Strengthen University Research Profiles (decision 301820) and the Photonics Research and Innovation Flagship (PREIN - decision 320165).
\end{acknowledgments}

\appendix \section{Switching off the LED}\label{ledoff}
Using these commands, one can switch off the LEDs in the case of low-light operations:

\noindent \texttt{i2 c set -y 0 0x15 0x00 0x00 i} \\
\texttt{i2 c set -y 0 0x15 0x01 0x00 i} \\
\texttt{i2 c set -y 0 0x15 0x02 0x00 i} \\
\texttt{i2 c set -y 0 0x15 0x03 0x00 i} \\
\texttt{i2 c set -y 0 0x15 0x04 0x00 i} \\

\noindent The different numbers are meant for different colours with \texttt{0x00} going for green all the way to \texttt{0x04} going for red. The latest part \texttt{0x00} indicates whether the LED is ON or OFF. 
It can be switched on by putting \texttt{0x01} instead of \texttt{0x00}. 
These operations can also be done by using the LightCrafter-Display-GUI software.

\section{DMD operation}\label{programDMD}

\subsection{SSH \& PuTTY}
Secure Shell (SSH) is a network protocol, which allows secure data transfer over an unsecured network. It is used for remote command execution and authentication. 
It provides security using encryption and cryptographic hash functions, and authenticates both the server and the user. The SSH protocol works over port 22. 
SSH enables a secure channel in a client-server architecture. 
In our case, the BeagleBone black acts as a server and the computer as a client.

PuTTY is a free (open-source) terminal emulator which supports SSH protocol and performs the remote command execution to display the patterns on the DMD.
The program can be downloaded on from the website: \href{http://www.putty.org/}{\textbf{www.putty.org}}.

\subsection{SFTP \& WinSCP}
Secure File Transfer Protocol (SFTP) is a network protocol to transfer files securely and reliably, with easy configuration. 
It uses the authentication and security provided by SSH.

WinSCP is a free (open-source) SFTP client supported by Windows. It provides an easy interface for connecting to a Linux based computer connected on the network and identified by an IP address, and exchange files between them. WinSCP was used to transfer files from the computer, where the patterns were generated, to the BeagleBone board connected with DMD. The program can be downloaded on from the website: \href{https://winscp.net/}{\textbf{www.winscp.net}}.

\subsection{Installation of image-viewer}
As the operating system is Debian, these series of commands will update the operating system and then install Feh image viewer.
\begin{itemize}
 	\item \texttt{sudo apt-get update} : updates the software database.
 	\item \texttt{sudo apt-get upgrade -y} : updates softwares already installed.
 	\item \texttt{sudo apt-get install feh} : installs Feh image viewer.
\end{itemize}
During this process, the device downloads the required files from the internet and installs them. This process may take 4-5 minutes. 

\section{Generating patterns using Matlab/Python}
As our DMD is 640$\times$360 pixels, we generate 2D black/white image having the same dimension as that of the DMD. 
Although, any programming language can be used for this purpose, we used Matlab. 
The mask images were saved in PNG format supported by Feh image viewer and transferred to the DMD. The programs are available as supplemental material.

\subsection{Single- and multiple-slit diffraction}
The single slit of thickness $d$ can be programmed using the equation
\begin{equation}
    P(x,y) = 
        \begin{cases} 
            1 & \text{if~} x_c-d/2 \leq x < x_c+d/2 \\
            0 & \text{otherwise}
        \end{cases}
\end{equation}
where $x_c=x_{max}/2$ is the center of the slit. For multiple-slit pattern, the value of $x_c$ is an array to obtain a desired number of slits.

\subsection{Fresnel zone plate}
The binary Fresnel zone plate geometry of focal length $f$ is constructed by alternating transparent and opaque zones, where the radius of the $n$th zone is given by 
\begin{equation}
    r_n^2 (r)=n\lambda \left( f+\frac{n\lambda}{r}\right)
\end{equation}
where $\lambda$ is the wavelength of the laser used, and $r=\sqrt{x^2+y^2}$ represent the radial coordinate system with coordinates $x,y$ in Cartesian coordinate system.

\subsection{Spatial modes}
For generating LG modes of OAM index $\ell$, the pattern for DMD can be calculated as
\begin{equation}
    P(x,y)=\frac{1+ sgn(\cos(kx-\ell\theta))}{2}
\end{equation}
where $\theta=\arctan(y,x)$. The parameter $k$ defines the grating period, which controls the separation from the central diffraction order in far field.

Note that with our DMD, HG modes can be generated using these holograms, harnessing the astigmatism of the DMD and Fourier transforming the generated LG modes.\cite{beijersbergen_astigmatic_1993}

\section{Diffraction grating analysis}
For single- and multple-slit diffraction, the theoretical expected curves can obtained from the relation \cite{hecht2002optics}
\begin{equation}
    I(\theta)=I(0)~\left( \frac{\sin(\beta)}{\beta}\right)^2 \left( \frac{\sin(N\alpha)}{\sin(\alpha)}\right)^2
\end{equation}
where $\beta=\pi d \sin (\theta)/\lambda$, $\alpha=\pi a \sin (\theta)/\lambda$, $d$ is the width of the slit, $a$ is the distance between the center of two slits, $\lambda$ is the wavelength of the laser, $N$ is the number of slits, and $\theta$ is the angle that the point of observation makes with the center of the grating structure. When comparing our measurements with curves obtained through this formula, we found a very good agreement. Specifically, we evaluated the correlation (see more details below) and obtained values of more than 94~\% for all results.

Alternatively, one can obtain the theoretically expected curves by using the split-step propagation method.\cite{poon2017engineering} 
All correlation values given in the main text are obtained using this method, which is also known as Fourier propagation method.

\section{Correlation function}
As a quantitative measure of the quality of our results, we compare them to the expected results from a simulation or direct calculation. 
We perform this analysis by evaluating the correlation (often called Pearson correlation coefficient), which is a measure of how close two variables are, i.e. correlated. 
The correlation $corr$ can be calculated from\cite{lee1988thirteen}
\begin{equation}
    corr = \frac{\sum{(M_i-\overline{M})(T_i-\overline{T})}}{\left[\sum{(M_i-\overline{M})^2}\sum{(T_i-\overline{T})^2}\right]^{1/2}},
\end{equation}
where $M$ and $T$ correspond to the measured and theoretical data, respectively, and $\overline{M},\overline{T}$ depicts their mean value.
Note, that there is also a direct Matlab command -\texttt{corrcoef}- to calculate it.  


\begin{thebibliography}{10}

\bibitem{weiner_femtosecond_2000}
A.~M. Weiner, ``Femtosecond pulse shaping using spatial light modulators,''
  Review of Scientific Instruments \textbf{71}, 1929--1960 (2000).

\bibitem{dholakia2008optical}
K.~Dholakia and W.~Lee, ``Optical trapping takes shape: the use of structured
  light fields,'' Advances in Atomic, Molecular, and Optical Physics
  \textbf{56}, 261--337 (2008).

\bibitem{Willner15}
A.~E. Willner, H.~Huang, Y.~Yan, Y.~Ren, N.~Ahmed, G.~Xie, C.~Bao, L.~Li,
  Y.~Cao, Z.~Zhao, J.~Wang, M.~P.~J. Lavery, M.~Tur, S.~Ramachandran, A.~F.
  Molisch, N.~Ashrafi, and S.~Ashrafi, ``Optical communications using orbital
  angular momentum beams,'' Advances in Optics and Photonics \textbf{7},
  66--106 (2015).

\bibitem{Rubinsztein-Dunlop2017}
H.~Rubinsztein-Dunlop, A.~Forbes, M.~V. Berry, M.~R. Dennis, D.~L. Andrews,
  M.~Mansuripur, C.~Denz, C.~Alpmann, P.~Banzer, T.~Bauer, E.~Karimi,
  L.~Marrucci, M.~Padgett, M.~Ritsch-Marte, N.~M. Litchinitser, N.~P. Bigelow,
  C.~Rosales-Guzm{\'{a}}n, A.~Belmonte, J.~P. Torres, T.~W. Neely, M.~Baker,
  R.~Gordon, A.~B. Stilgoe, J.~Romero, A.~G. White, R.~Fickler, A.~E. Willner,
  G.~Xie, B.~McMorran, and A.~M. Weiner, ``Roadmap on structured light,''
  Journal of Optics \textbf{19}, 013001 (2017).

\bibitem{erhard2018twisted}
M.~Erhard, R.~Fickler, M.~Krenn, and A.~Zeilinger, ``Twisted photons: new
  quantum perspectives in high dimensions,'' Light: Science \& Applications
  \textbf{7}, 17146--17146 (2018).

\bibitem{forbes2016creation}
A.~Forbes, A.~Dudley, and M.~McLaren, ``Creation and detection of optical modes
  with spatial light modulators,'' Advances in Optics and Photonics \textbf{8},
  200--227 (2016).

\bibitem{rosales-guzman_how_2017}
C.~Rosales-Guzmán and A.~Forbes, \emph{How to shape light with spatial light
  modulators} (SPIE Digital Library, 2017).

\bibitem{padgett1996experiment}
M.~Padgett, J.~Arlt, N.~Simpson, and L.~Allen, ``An experiment to observe the
  intensity and phase structure of {Laguerre-Gaussian} laser modes,'' American
  Journal of Physics \textbf{64}, 77--82 (1996).

\bibitem{boruah2009dynamic}
B.~Boruah, ``Dynamic manipulation of a laser beam using a liquid crystal
  spatial light modulator,'' American Journal of Physics \textbf{77}, 331--336
  (2009).

\bibitem{jones2016poincare}
J.~A. Jones, A.~J. D’Addario, B.~L. Rojec, G.~Milione, and E.~J. Galvez,
  ``The {P}oincar{\'e}-sphere approach to polarization: Formalism and new labs
  with {P}oincar{\'e} beams,'' American Journal of Physics \textbf{84},
  822--835 (2016).

\bibitem{scholes_structured_2019}
S.~Scholes, R.~Kara, J.~Pinnell, V.~Rodríguez-Fajardo, and A.~Forbes,
  ``Structured light with digital micromirror devices: a guide to best
  practice,'' Optical Engineering \textbf{59}, 041202 (2019).

\bibitem{huang2012low}
D.~Huang, H.~Timmers, A.~Roberts, N.~Shivaram, and A.~S. Sandhu, ``A low-cost
  spatial light modulator for use in undergraduate and graduate optics labs,''
  American Journal of Physics \textbf{80}, 211 (2012).

\bibitem{Dmdappl1994}
J.~B. Sampsell, ``Digital micromirror device and its application to projection
  displays,'' Journal of Vacuum Science \& Technology B \textbf{12}, 3242--3246
  (1994).

\bibitem{beaglebone}
B.~Black, ``{BeagleBone Black},''  (2020). Https://beagleboard.org/black.

\bibitem{dlp2000EVM}
T.~Instruments, ``{DLP® LightCrafter™ Display 2000 Evaluation Module},''
  (2020). Https://www.ti.com/tool/DLPDLCR2000EVM.

\bibitem{dlpprojector}
M.~LLC, ``Building an embedded {DLP} projector for \$99 and a beaglebone,''
  (2020).
  Https://www.electronicdesign.com/technologies/embedded-revolution/article/21805540/building-an-embedded-dlp-projector-for-99-and-a-beaglebone.

\bibitem{galvez2006gaussian}
E.~J. Galvez, ``Gaussian beams in the optics course,'' American Journal of
  Physics \textbf{74}, 355--361 (2006).

\bibitem{dlp2000}
T.~Instruments, ``{DLP2000},''  (2020). Https://www.ti.com/product/DLP2000.

\bibitem{dlppicooverview}
T.~Instruments, ``{DLP Pico Display overview},''  (2020).
  Https://www.ti.com/dlp-chip/display-and-projection/pico-chipsets/overview.html.

\bibitem{mirhosseini2013rapid}
M.~Mirhosseini, O.~S. Magana-Loaiza, C.~Chen, B.~Rodenburg, M.~Malik, and R.~W.
  Boyd, ``Rapid generation of light beams carrying orbital angular momentum,''
  Optics Express \textbf{21}, 30196--30203 (2013).

\bibitem{lee1979binary}
W.-H. Lee, ``Binary computer-generated holograms,'' Applied Optics \textbf{18},
  3661--3669 (1979).

\bibitem{poon2017engineering}
T.-C. Poon and T.~Kim, \emph{Engineering Optics with MATLAB} (World Scientific
  Publishing Company, 2017).

\bibitem{lee1988thirteen}
J.~Lee~Rodgers and W.~A. Nicewander, ``Thirteen ways to look at the correlation
  coefficient,'' The American Statistician \textbf{42}, 59--66 (1988).

\bibitem{goodman2005introduction}
J.~W. Goodman, \emph{Introduction to Fourier optics} (Roberts and Company
  Publishers, 2005).

\bibitem{hecht2002optics}
E.~Hecht, \emph{Optics} (Pearson Education, 2002), 5th ed.

\bibitem{jesacher2007wavefront}
A.~Jesacher, A.~Schwaighofer, S.~F{\"u}rhapter, C.~Maurer, S.~Bernet, and
  M.~Ritsch-Marte, ``Wavefront correction of spatial light modulators using an
  optical vortex image,'' Optics Express \textbf{15}, 5801--5808 (2007).

\bibitem{allen1992orbital}
L.~Allen, M.~W. Beijersbergen, R.~Spreeuw, and J.~Woerdman, ``Orbital angular
  momentum of light and the transformation of {Laguerre-Gaussian} laser
  modes,'' Physical Review A \textbf{45}, 8185 (1992).

\bibitem{Fickler2016}
R.~Fickler, G.~Campbell, B.~Buchler, P.~K. Lam, and A.~Zeilinger, ``Quantum
  entanglement of angular momentum states with quantum numbers up to 10,010,''
  Proceedings of the National Academy of Sciences \textbf{113}, 13642 (2016).

\bibitem{carpentier_making_2008}
A.~V. Carpentier, H.~Michinel, J.~R. Salgueiro, and D.~Olivieri, ``Making
  optical vortices with computer-generated holograms,'' American Journal of
  Physics \textbf{76}, 916 (2008).

\bibitem{Lerner:12}
V.~Lerner, D.~Shwa, Y.~Drori, and N.~Katz, ``Shaping {Laguerre-Gaussian} laser
  modes with binary gratings using a digital micromirror device,'' Optics
  Letters \textbf{37}, 4826--4828 (2012).

\bibitem{beijersbergen_astigmatic_1993}
M.~Beijersbergen, L.~Allen, H.~van~der Veen, and J.~Woerdman, ``Astigmatic
  laser mode converters and transfer of orbital angular momentum,'' Optics
  Communications \textbf{96}, 123 (1993).

\bibitem{curtis2003structure}
J.~E. Curtis and D.~G. Grier, ``Structure of optical vortices,'' Physical
  Review Letters \textbf{90}, 133901 (2003).

\end{thebibliography}
\end{document}